\documentstyle[psfig]{lamuphys}
\makeatletter
\let\chapter\hid@chapter
\makeatother

\begin{document}
\pagenumbering{arabic}
\title{Properties of High Redshift Cluster Ellipticals}

\author{Mark Dickinson}

\institute{Space Telescope Science Institute, Baltimore MD 21218, USA}

\maketitle

\begin{abstract}
Cluster ellipticals are often thought to be among the oldest galaxies
in the universe, with the bulk of their stellar mass formed at early
cosmic epochs.  I review recent observations of color evolution in 
early--type cluster galaxies at high redshift, which show remarkably
little change in the color--magnitude relation out to $z \approx 1$.
Spectra of elliptical galaxies from $1.15 < z < 1.41$ demonstrate the
presence of a dominant old stellar population even at these large 
lookback times, although there is some evidence for a tail of later
star formation.  The Kormendy relation, a photometric/morphological
scaling law for elliptical galaxies, is used to extend fundamental plane
investigations to $z = 1.2$ and to test for luminosity evolution.
While the evidence, overall, is consistent with simple and mild passive
evolution, I review some caveats and briefly consider the data in the 
light of hierarchical galaxy formation models.
\end{abstract}

\section{Color evolution at $z <$ 1}

\begin{figure}
\centerline{\psfig{figure=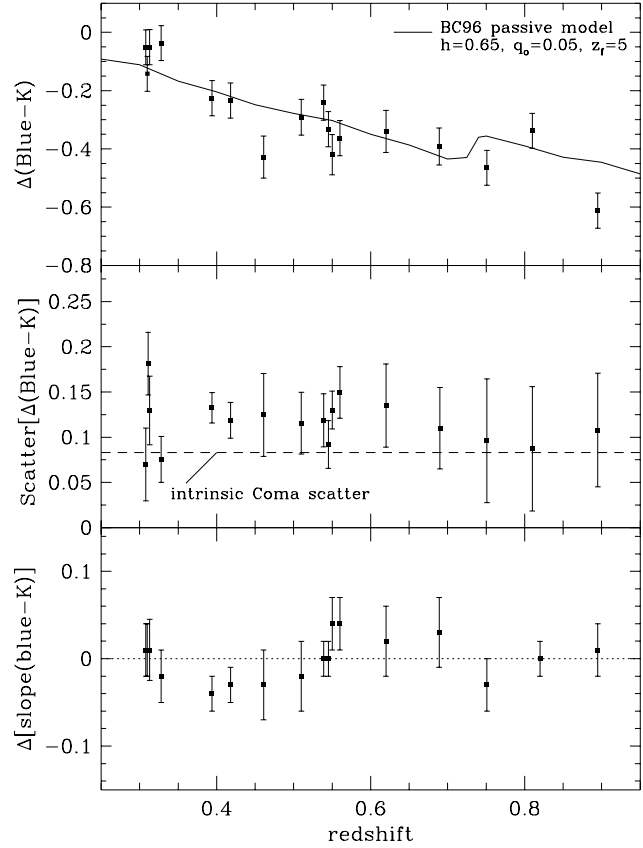,height=5in}}
\caption{
Color evolution of early--type cluster galaxies out to $z=0.9$,
from Stanford, Eisenhardt \& Dickinson 1997.  The data consist of 
optical and infrared photometry of early--type galaxies selected
from WFPC2 images of high redshift clusters.  The ``blue'' band is 
chosen for each cluster to remain approximately fixed at the rest--frame
$U$--band, while the $K$ band remains fixed in the observer's frame.
{\it Top:} differences in mean $blue - K$ colors relative to the same
rest--frame colors in the Coma cluster.  ``No evolution'' would be therefore
be a horizontal line at $\Delta(blue - K) = 0$.  
{\it Middle:}~intrinsic scatter in early--type galaxy colors, after removing
the mean slope of the color--magnitude relation and the component of
scatter due to photometric errors.  {\it Bottom:} differences
in the slope of the $(blue - K)$ vs. $K$ color--magnitude relation, 
relative to the slope measured for Coma.
}
\end{figure}

Rich clusters remain the most ``efficient'' place to find and study
high redshift elliptical galaxies -- they are abundant in such environments,
even at $z > 0.5$, and the richness of the cluster ensures that a statistical
analysis of their colors or image properties (in the absence of extensive
spectroscopy) should not be too strongly
biased by field galaxy ``interlopers.''
Until recently, studies of distant cluster ellipticals 
were predicated on selecting them by color or spectral properties,
e.g. by analyzing the colors of the ``red envelope'' of the galaxy
distribution and assuming that any changes reflect the evolution
of the early--type galaxy population (e.g. Arag\'on--Salamanca
{\it et al.} 1993).   While sensible, this approach requires
a reliable statistical means of identifying and measuring properties
of galaxies in a one--sided distribution -- e.g. the ``reddest objects''
in a given cluster -- and risks missing important results if, for example,
some {\it bona fide} ellipticals are substantially bluer than their reddest
counterparts.

Now, however, a very extensive collection of WFPC2 images of high redshift
clusters has amassed in the HST archive, enabling the armchair astronomer
to ponder hundreds of high redshift ellipticals at will, selecting them
on the basis of image morphology rather than color.   
The ``MORPHS'' collaboration (Dressler, Oemler, Ellis \& many co--workers)
have been responsible for collecting much of this data, and have recently
analyzed the properties of early--type galaxies in three clusters at
$z \approx 0.55$ (Ellis {\it et al.} 1997).  In particular, they have
examined the scatter in the elliptical galaxy color--magnitude (c--m)
relation.  Bower {\it et al.} (1992) used the very small scatter in the 
c--m relation for Coma and Virgo ellipticals to argue that these galaxies
have had highly synchronous evolutionary histories:  they either formed
at very large redshifts, or with nearly identical subsequent star formation
histories, or both.   Ellis {\it et al.} find the same to be true
in the three clusters they examined at $z \approx 0.55$, reinforcing the
evidence for synchronized evolution.

Adam Stanford, Peter Eisenhardt and I have collected 5--band optical/IR imaging
of a large sample of distant galaxy clusters (46 to date from $0 < z < 0.9$),
taking care to match both the field of view (in Mpc) and the
rest--frame limiting luminosity of the data for each cluster in order to
provide a uniform data set. 
The optical/CCD observations were taken through two filters
which are matched to the cluster redshift in order to approximately sample
the rest--frame $U$ and $V$ regions of the spectrum, while the near--IR
$JHK$ data provide a long wavelength baseline for measuring colors
and ensure that galaxies may be selected uniformly by the luminosities 
of their old stellar populations, even at $z = 1$.  

Returning from the telescopes to our armchairs, we have used archival WFPC2 
data to identify early--type galaxies, independently from their colors, 
in 19 clusters, and to study their photometric evolution (Stanford, 
Eisenhardt \& Dickinson 1997).  We have used new
wide-field imaging of Coma cluster in the $UBVRIzJHK$ bands (Eisenhardt
{\it et al.} 1997) to provide a present--day reference sample, ensuring that 
we can compare the properties of the distant cluster galaxies to data on 
nearby ellipticals at the same rest--frame wavelengths with only minimal 
differential k--corrections.  Figure 1 summarizes this work, showing the 
evolution of the galaxy colors, and of the scatter and slope of the 
color--magnitude relation, out to $z=0.9$.  The key results are:
\begin{itemize}
\item 
The mean color of early--type cluster galaxies becomes
gradually and monotonically bluer toward higher redshifts, in a fashion
broadly consistent with simple passive evolution of the stellar populations.
\item
The scatter of galaxy colors around the mean color--magnitude relation
remains small and nearly constant with redshift out to $z=0.9$.  
\item
The slope of the color--magnitude relation remains unchanged from
$z=0.9$ to the present.   This strongly supports the hypothesis that
the c--m slope is primary due to differences in the mean metallicity of
elliptical galaxies as a function of luminosity/mass, and is not a consequence
of differing ages (Kodama \& Arimoto 1997;  see also contribution of Arimoto 
to this volume).
\end{itemize}

The small and constant scatter in the galaxy colors 
is remarkable and somewhat difficult to explain.  
In particular, if ellipticals in all clusters were a completely coeval 
population, and if the small but non--zero scatter observed in their
colors at $z=0$ (Bower {\it et al.} 1992) were due to age variations,
then one would naively expect the scatter to increase at higher redshift.  
Although the mean scatter in the distant clusters is slightly higher
than that measured for Coma, no other strong trend with redshift is
observed.  It may be that the intrinsic scatter at $z=0$ is
due, in part, to metallicity variations at a fixed mass rather than
to age differences.  This would set a ``floor'' value to the scatter, 
suppressing the expected decrease with decreasing redshift.  
In this case, however, intrinsic age variations must be even smaller 
than the values inferred directly from the color scatter.
Alternatively, small episodes of later star formation due to mergers, 
etc. may ``re--inflate'' the scatter in a more--or--less continuous fashion,
but again the amount of late star formation is strongly constrained by
the small amplitude of the measured scatter.  The 0.11 magnitude scatter
observed for the most distant cluster in our sample, GHO~1603+4313 at 
$z=0.895$, limits small (e.g. $\leq 10\%$ by mass) star formation
episodes to have occurred no less than 2~Gyr prior to the epoch of
observation.  

\section{Spectroscopic characteristics of ellipticals at $z >$ 1}

The lack of dramatic change in the color properties of early--type cluster 
galaxies out to $z=0.9$ suggests that, by that redshift, we have not 
closely approached the era in which those galaxies formed the
bulk of their stars.  This therefore encourages us to extend the search 
for distant clusters and cluster ellipticals to higher redshifts.  
Peter Eisenhardt and I have been studying the environments of distant radio
galaxies in order to search for distant galaxy clusters (cf. Dickinson
1995, 1997a, 1997b).  Spinrad, Dey, Stern, LeF\`evre and I have obtained
extensive spectroscopy of one such cluster, that around 3C324 at 
$z=1.206$.   Using deep WFPC2 imaging and infrared photometry, we have 
identified elliptical galaxies in this cluster, as well 
as several in a ``foreground'' structure at $z=1.15$ and one background 
elliptical at $z=1.41$.  

\begin{figure}
\centerline{\psfig{figure=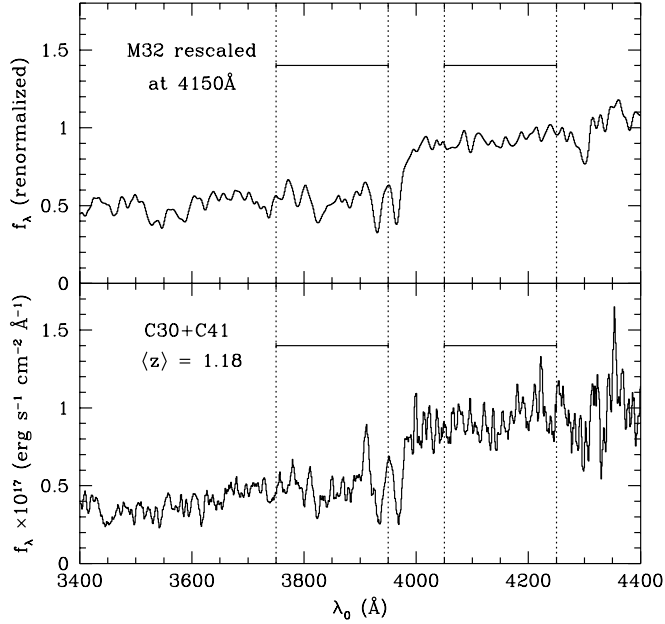,height=3.5in}}
\caption{Coadded spectra in the rest--frame 4000\AA\ region of two elliptical 
galaxies at $z=1.15$ and $1.21$ in the field of 3C324 (bottom) compared to
that of M32 (top).  A very prominent 4000\AA\ break and CaII H+K lines are
visible;  the spectral regions used to define the D4000 break amplitude 
are marked.}
\end{figure}

\begin{figure}
\centerline{\psfig{figure=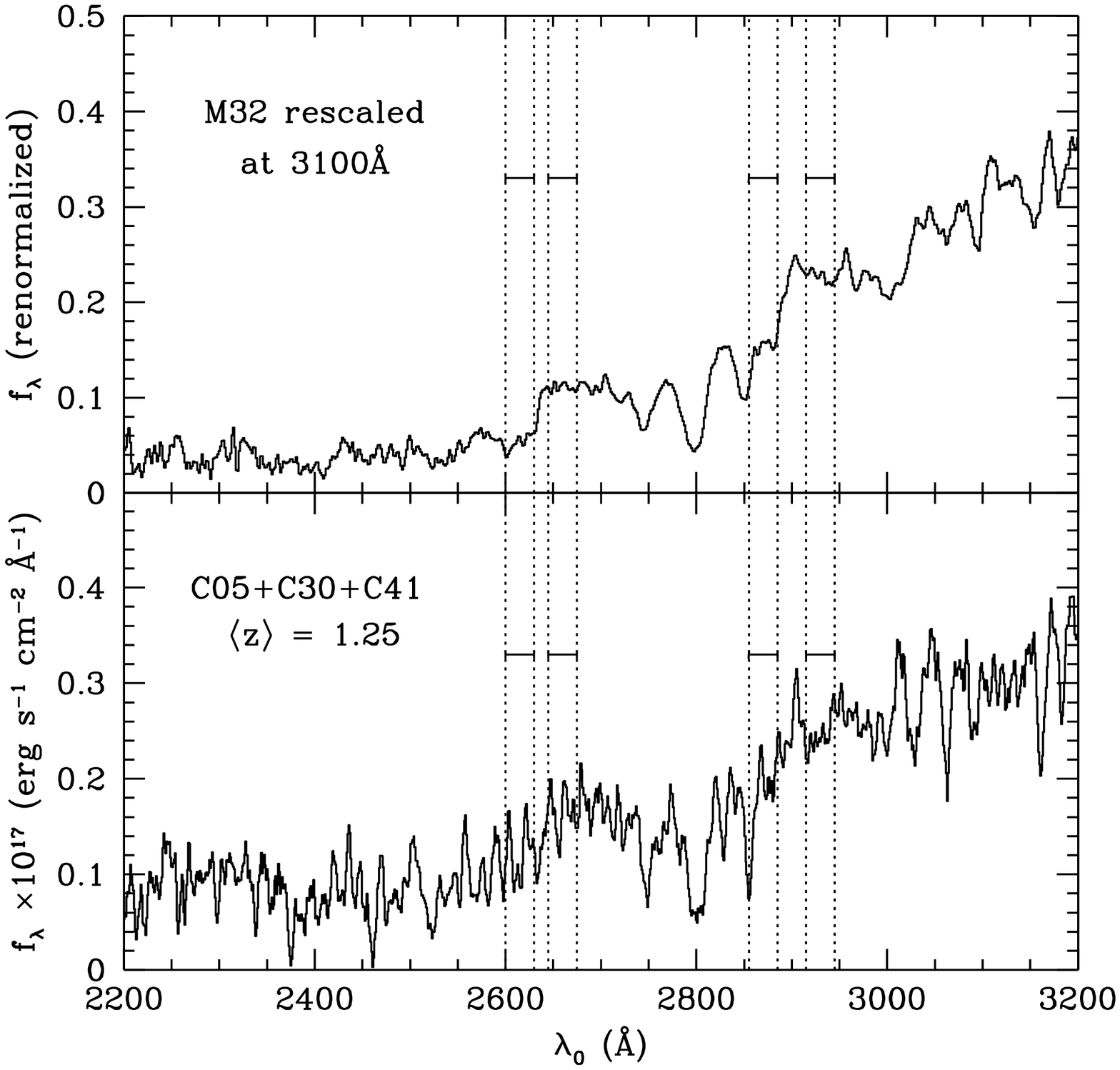,height=3.5in}}
\caption{Coadded near--UV rest--frame spectra of three elliptical 
galaxies at $z=1.15$ to $1.41$ in the field of 3C324 (bottom) compared to
an IUE spectrum of of M32 (top).  The strong MgII 2800\AA\ absorption feature 
is clearly seen in the high redshift objects, flanked by MgI 2852\AA\ and the 
absorption blend (primarily Fe+Cr) at 2746\AA.  The dotted lines mark the
2640\AA\ and 2900\AA\ break features defined by Spinrad {\it et al.} 1997.}
\end{figure}

Figures 2 and 3 present coadded Keck/LRIS spectra of three of these galaxies, 
showing the rest--frame optical and near--UV regions and identifying
some of the prominent spectral breaks which are useful as diagnostics 
of the stellar populations.  The $z \approx 1.2$ galaxies exhibit a strong
4000\AA\ break and CaII~H+K lines, features prominent in the spectra
of old stellar populations today.  The near--UV spectra are strikingly
similar to that of M32, although they are somewhat bluer (more flux
at $\lambda_0 < 2600$\AA\ relative to that at $\lambda_0 \approx 3100$\AA)
and the discontinuities at 2640\AA\ and 2900\AA\ have somewhat
smaller amplitudes.

Dunlop {\it et al.} (1996) have recently presented spectra of the faint, 
red radio galaxy 53W091 at even larger redshift, $z = 1.552$, which exhibits
very similar UV spectral breaks at 2640\AA\ and 2900\AA.\  Spinrad 
{\it et al.} (1997) have analyzed the 53W091 data, making extensive 
comparisons to the ultraviolet properties of stars in the IUE spectral 
atlas, to nearby elliptical galaxies, and to population synthesis 
models -- the reader is referred to 
that paper for a more thorough discussion, as well as to 
Charlot, Worthey \& Bressan (1996) for an excellent review 
of potential pitfalls in the use of the evolutionary models.  

\begin{figure}
\centerline{\psfig{figure=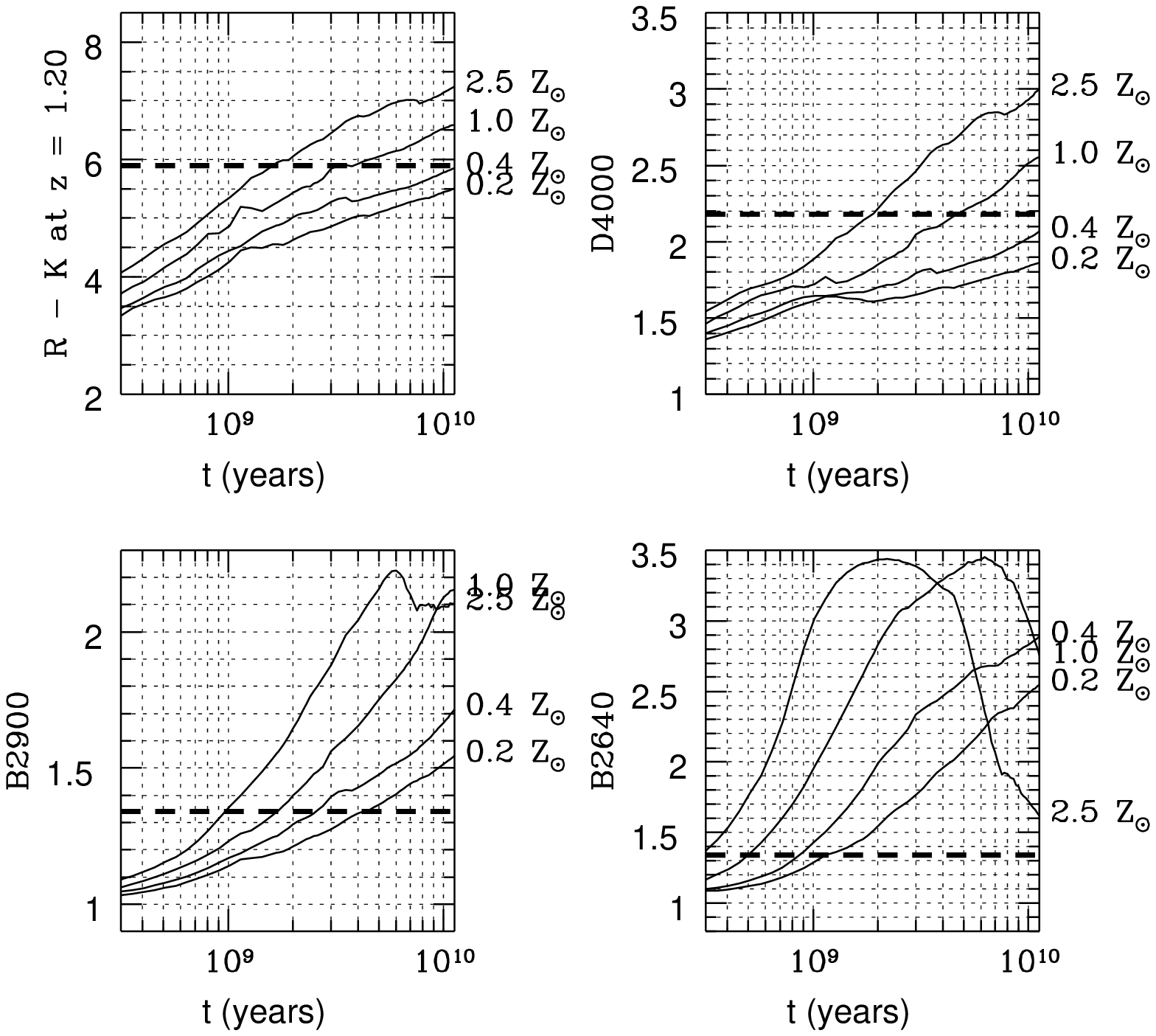,width=5.4in}}
\caption{Simple models of passive evolution of a stellar population
formed in a $10^7$ year burst with a Salpeter IMF, computed using the 
code of Bruzual \& Charlot (1996) for various metallicities.  The panels 
show the expected amplitudes of three spectral ``breaks'' and the
$R-K$ broad band color.  The $R$--band measures rest--frame light at
$\approx 3150$\AA\ for $z=1.2$, and thus is primarily sensitive to
evolution at UV wavelengths intermediate between those measured by
the 4000\AA\ and 2900\AA\ breaks.  The horizontal dashed lines indicate
the values measured for the $z > 1$ ellipticals discussed in the text.}
\end{figure}

Here, we restrict our comparison to a single set of evolutionary models, 
those of Bruzual \& Charlot (1996), which utilize model stellar atmospheres 
and evolutionary tracks to investigate the spectrophotometric evolution of 
stellar populations with various metallicities.  Figure~4 shows the 
dependence of four indices (the three spectral breaks plus the $R-K$ color) 
on age and metallicity for simple Salpeter IMF models ($10^7$~yr burst 
followed by purely passive evolution).  The horizontal dashed lines mark the
index values for the 3C324 galaxies.   All four indices exhibit the familiar
age--metallicity degeneracy (e.g. Worthey 1994) -- redder colors or larger
break value may be matched either by older or more metal rich stellar 
populations.  Spinrad {\it et al.} review evidence that an approximately solar
metallicity is an appropriate average value for giant ellipticals;  we 
leave this problem unresolved here, but note that by studying the spectral 
properties of galaxies over a large range of cosmic time (rather than at 
a single, large redshift) we may eventually hope to break this degeneracy,
as different rates of evolution would be expected for different
metallicities.

Inspecting the four panels, one finds that the ``age'' derived from 
the various indices becomes ``younger'' the farther into the
UV we look.  Moving from red to blue indices (D4000, $R-K$, B2900
and B2640), and taking the solar metallicity model for reference, 
the ages derived for the single--burst model are approximately 
4.5, 3.5, 1.7 and 0.5 Gyr, respectively.  This suggests that the 
single, short--burst star formation history is not strictly correct, 
and that younger starlight plus an increasing role as one moves farther 
into the rest--frame UV.  Models with more protracted star formation 
histories (e.g. exponentially decaying) or with later bursts can be 
constructed which reasonably match the data.  However, the large 
4000\AA\ break amplitude and $R-K$ color strongly indicate that
the bulk of the stars in these galaxies must have formed {\it at least} 
2~Gyr prior to $z=1.2$, even for populations with super--solar
metallicities.  For an open, 13~Gyr--old universe  ($H_0 = 75$,
$q_0 = 0$) this places the 
formation redshift for the bulk of the stellar mass very conservatively 
at $z > 2.3$, and most probably higher.  For a closed universe, 
formation can be pushed back to almost arbitrarily large redshifts.

\section{Scaling relations to $z =$ 1.2}

Some of the most exciting data shown at this meeting have been the
fundamental plane and Mg--$\sigma$ observations at $0.1 < z < 0.6$, 
presented in the contributions of Bender, Franx, J\o rgensen,
Pahre, van Dokkum, and Ziegler to these proceedings (cf. also 
van Dokkum \& Franx 1996 and Kelson {\it et al.} 1997).  The conclusions 
of these various studies have been strikingly uniform:\ that the 
mass--to--light ratios of early--type galaxies in distant clusters 
have evolved very mildly, in a manner approximately consistent with 
expectations from simple, passive evolution of their stellar populations.  

\begin{figure}
\centerline{\psfig{figure=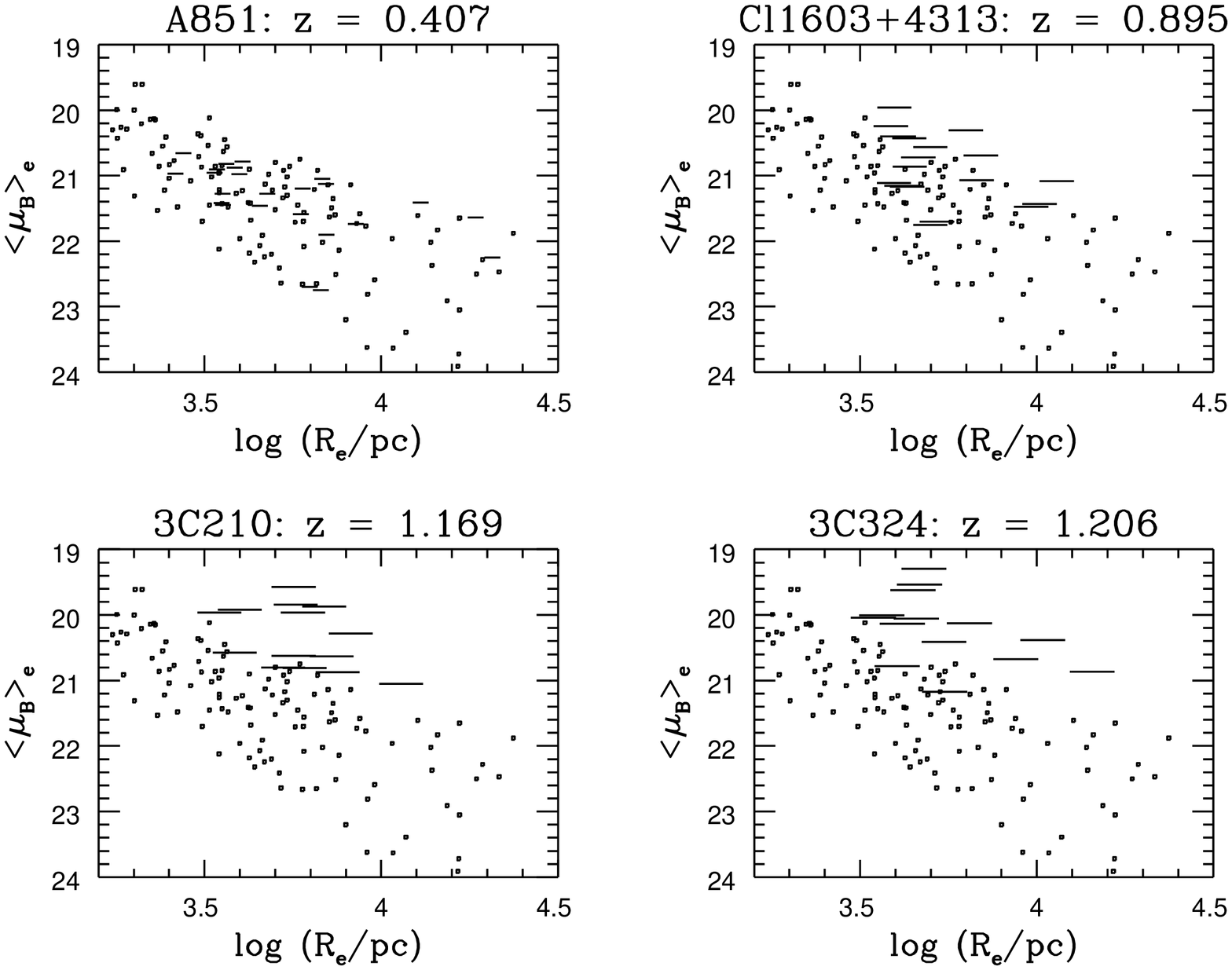,width=5.4in,angle=-90}}
\caption{The Kormendy relation for four high redshift galaxy clusters 
observed with HST.  The small points are a sample of local cluster
galaxies observed by J\o rgensen {\it et al.} (1995).  The horizontal
lines represent individual high redshift ellipticals, and connect the
$R_e$ values for $q_0 = 0$ and $q_0 = 0.5$ ($H_0 = 50$ assumed throughout).}
\end{figure}

\begin{figure}
\centerline{\psfig{figure=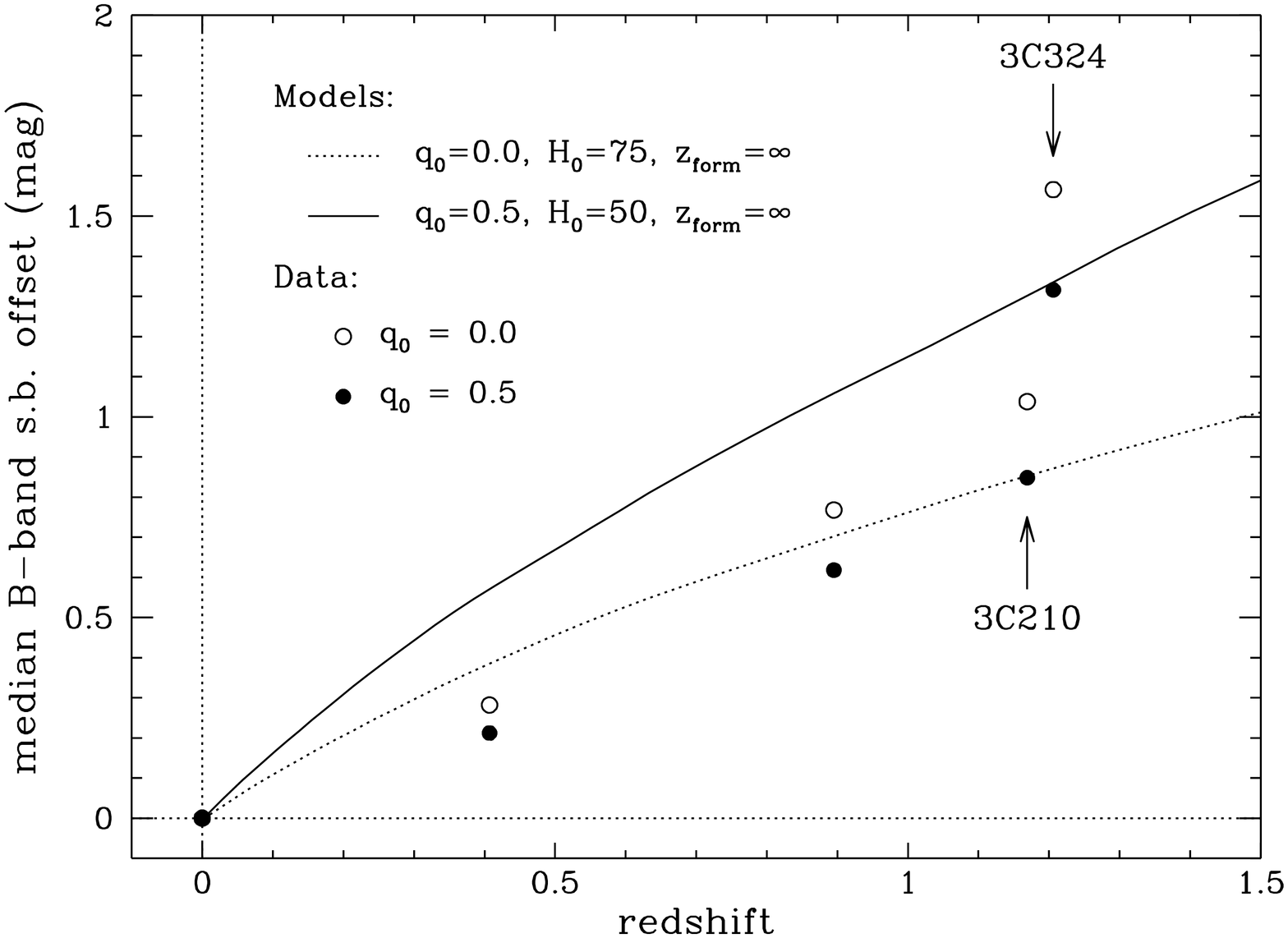,width=4.5in,angle=-90}}
\caption{Median rest--frame $B$--band surface brightness offsets of
distant cluster ellipticals relative to the local Kormendy relation.}
\end{figure}

At present it is still difficult to extend this work to much larger 
redshifts, even with 10m--class telescopes, because the galaxies become 
extremely faint and the spectral features of interest for measuring 
velocity dispersions or Mg indices shift into the near infrared where 
they are horrendously impacted by OH night sky emission.
In the absence of high--S/N, high dispersion spectroscopy 
at $z > 0.8$, we may at least take advantage of purely
photometric/morphological projections of the fundamental plane which
allow us to push studies of galaxy scaling relations to higher redshifts.
For example, the Kormendy relation is the projection of the fundamental
plane onto the size vs. surface brightness axes, and is thus highly suitable
for study using HST WFPC2 images alone.  Dickinson (1995), 
Pahre {\it et al.} (1996), Barrientos {\it et al.} (1996), and 
Schade {\it et al.} (1996 and this volume) have all employed variants on
this technique.  The method has the advantage of requiring only 
high--resolution imaging data which is readily obtained for a large
number of cluster galaxies per WFPC2 field, but it lacks the precision of
the direct fundamental plane measurements (due to the larger scatter
in the Kormendy relation);  moreover without the kinematic data offered
by spectroscopy one cannot directly connect the observables to evolution
in the mass--to--light ratio of distant galaxies.

Figure 5 presents an update on the data shown in Dickinson (1995), 
demonstrating the Kormendy relation in the rest--frame $B$--band for four 
high redshift clusters, including two at $z \approx 1.2$.  
As one moves to higher redshifts, the cluster galaxies
lie increasingly ``above'' the locus of low--redshift ellipticals, 
indicating either higher rest--frame surface brightnesses, larger 
effective radii, or some combination thereof.  It is simplest, 
but not definitive, to interpret this as a manifestation of passive 
luminosity evolution.   Figure~6 plots the median offsets of the 
Kormendy relation for each cluster from the local values, and compares 
them with expectations from passively evolving Bruzual--Charlot
(1996) models for two different cosmologies, each assuming present--day
ages for galaxies (and the universe as a whole) of 13~Gyr.
The two $z \approx 1.2$ clusters seem to differ significantly from one
another, but
it is important to note that the 3C324 data was obtained with the
WFPC2 F702W filter, which samples the cluster galaxy light at a rest
frame of $\lambda_0 \approx 3200$\AA, requiring a large (and uncertain)
k--correction to rest--frame $B$.\  3C210 is at slightly lower redshift,
and was observed with F814W (rest frame 3700\AA), resulting in a much
smaller k--correction.  Overall, the implied luminosity evolution
out to $z \approx 1$ is quite small compared to the stellar population
models (see also van Dokkum \& Franx 1996), and more consistent with the
open universe models than the closed ones.

\section{Discussion}

Most of the effort directed toward studying distant elliptical galaxies 
has been aimed either at verifying the gospel of monolithic formation
and passive evolution, or at falsifying this scenario as a heinous fiction 
which flies in the face of the cold logic of hierarchical merging.
While most of the evidence presented above would seem consistent with 
the idea that early--type cluster galaxies have primarily undergone
quiescent and passive evolution from $z \approx 1$ to the present, it
is important to remember the distinction between the galaxies which
we {\it observe} as ellipticals at high redshift and the exact progenitors
of today's elliptical galaxies.  While avoiding color--dependent selection 
effects, morphological selection of high redshift, early--type galaxies 
may itself introduce biases.  By studying things that {\it look} like 
ellipticals, one is not necessarily tracing the past history of the actual 
ancestors to {\it all} of today's elliptical galaxies.  In particular, 
if substantial merging has taken place to form the elliptical population, 
then the objects which are recognizably elliptical galaxies at any redshift 
may be only the oldest descendents (at that epoch) of that merging process, 
while more recent (or still--to--be) mergers may not enter into a 
morphologically--selected catalog.   We may only be studying the most
dynamically evolved galaxies in each cluster at each redshift, and thus
those for which ``recent'' merger--induced star formation is pushed back
to earlier and earlier cosmic times.  Moreover, we have not attempted to 
separate ellipticals from S0 galaxies in our data.  Recently, 
Dressler \& Smail (1997) have suggested that there is a marked absence 
of S0 galaxies in distant clusters imaged by WFPC2.  The implication is 
that the Butcher--Oemler effect may be, in part, a consequence of disk 
galaxies being transformed into S0s.   Depending on the extent
of star formation associated with this transformational process, this could 
provide a mechanism for ``re--inflating'' the scatter in the combined E/S0
color--magnitude relation at late times, thus giving the impression of
no evolution in that scatter.

Regardless, it seems to be clear that a substantial population of early--type
galaxies was present in rich clusters out to $z \approx 1$ and beyond, and that
the bulk of their stars must have formed at substantially larger redshifts.
Whether {\it all} ellipticals formed at such early times is less 
certain, but there is little evidence from the cluster data to
indicate that it is not so.   It is important to emphasize that the data 
considered here has primarily been for elliptical galaxies in rich clusters,
and that field galaxies might have had rather different evolutionary histories.
Kauffmann {\it et al.} (1996) have used data from the Canada--France
Redshift Survey to argue for strong number--density evolution in 
the population of early--type field galaxies since $z \approx 1$.  Their
argument is basically that the apparent {\it lack} of strong evolution
in the luminosity function of early--type (or at least red) galaxies in 
the CFRS (Lilly {\it et al.} 1995) contradicts expectations for passive 
stellar evolution (see figure 6), and thus must be counterbalanced 
by extensive number--density evolution.  For cluster ellipticals, 
the hierarchical merging models (e.g. reviewed by White in this volume;
cf. also Baugh {\it et al.} 1996) again predict that while many 
or most of their {\it stars} formed at $z > 1$, the galaxies themselves
assembled late, with $\sim$70\% forming after $z = 1$.  This is difficult
to test from the existing cluster data -- the ellipticals already present 
in clusters at $z \approx 1$ cannot easily tell us how many more will join
them by $z = 0$.  Moreover, the richest clusters, which are generally
those which we observe at high redshift, are quite probably
the most unusual and overdense environments at those epochs, and thus are
the places where the galaxy merging history is ``pushed back'' to the
earliest cosmic times.  The future of this work lies in the investigation
of clusters at still larger redshifts, but equally importantly in the
analysis of early--type galaxies across a broad range of environments at 
$z < 1$, from rich clusters through poorer systems and groups and into the
field.

\vspace{0.5cm}

\noindent {\bf Acknowledgments:\ } I would like to thank my collaborators, 
particularly Adam Stanford,
Peter Eisenhardt, Hy Spinrad, Arjun Dey, Dan Stern, and Olivier LeF\`evre,
for allowing me to present material here in advance of publication.  
My thanks also to Stephane Charlot for assistance with the population
synthesis models, to the conference organizers for their hospitality and
financial support, and to the editor of these proceedings for his
patience.


\begin{thebibliography}

\bibitem{}{}{} Arag\'on--Salamanca, A., Ellis, R.S., Couch, W.J., and
Carter, D., 1993, MNRAS, 248, 128.

\bibitem{}{}{} Barrientos, F., Schade, D., and L\'opez--Cruz, O., 1996,
Ap.J., 460, 89.

\bibitem{}{}{} Baugh, C.M., Cole, S., and Frenk, C.S., 1996, MNRAS, 283, 1361.

\bibitem{}{}{} Bower, R.G., Lucey, J.R., and Ellis, R.S., 1992,
MNRAS, 254, 601.

\bibitem{}{}{} Bruzual, G., and Charlot, S. 1996, private communication.

\bibitem{}{}{} Charlot, S., Worthey, G., and Bressan, A., 1996, 
Ap.J., 457, 625.

\bibitem{}{}{} Dickinson, M. 1995, in {\it Fresh Views of Elliptical Galaxies,}
eds. A.~Buzzoni, A.~Renzini, \& A.~Serrano, ASP, San Francisco, p. 283.

\bibitem{}{}{} Dickinson, M., 1997a, in {\it The Early Universe with the VLT,}
ed. J. Bergeron, Springer--Verlag, Berlin, p. 274.

\bibitem{}{}{} Dickinson, M., 1997b, in {\it HST and the High Redshift 
Universe,}
eds. N. Tanvir, A. Arag\'on--Salamanca, and J.V. Wall, World Scientific,
in press.

\bibitem{}{}{} Dressler, A., and Smail, I., 1997, in 
{\it HST and the High Redshift Universe,} eds. N. Tanvir, 
A. Arag\'on--Salamanca, and J.V. Wall, World Scientific, in press.

\bibitem{}{}{} Dunlop, J., Peacock, J., Spinrad, H., Dey, A., Jimenez, R.,
Stern, D., and Windhorst, R., 1996, Nature, 381, 581.

\bibitem{}{}{} Eisenhardt, P.R.M., Stanford, S.A., Dickinson, M., and
de Propris, R. 1997, in prep.

\bibitem{}{}{} Ellis, R.S., Smail, I., Dressler, A., Couch, W.J., Oemler, A.,
Butcher, H., and Sharples, R.M., 1997, Ap.J, in press.

\bibitem{}{}{} Kauffmann, G., Charlot, S. and White, S.D.M., 1996, 
MNRAS, 283, 117.

\bibitem{}{}{} Kelson, D.D., van Dokkum, P., Franx, M., Illingworth, G.D.,
and Fabricant, D. 1997, Ap.J., in press.

\bibitem{}{}{} Lilly, S.J., Tresse, L., Hammer, F., Crampton, D., and 
LeF\`evre, O., 1995, ApJ, 455, 108L.

\bibitem{}{}{} Pahre, M.A., Djorgovski, S.G., and DeCarvalho, R.R. 1996,
ApJ, 456, L79.

\bibitem{}{}{} Schade, D., Barrientos, L.F., and L\'opez--Cruz, O., 1997,
Ap.J., in press.

\bibitem{}{}{} Spinrad, H., Dey, A., Stern, D., Dunlop, J., 
Peacock, J., Jimenez, R., and Windhorst, R. 1997, Ap.J., in press.

\bibitem{}{}{} Stanford, S.A., Eisenhardt, P.R.M., and Dickinson, M. 1997, 
ApJ, submitted.

\bibitem{}{}{} J\o rgensen, I., Franx, M., and Kjaergaard, P. 1995, MNRAS, 
273, 1097.

\bibitem{}{}{} Kodama, T., and Arimoto, N. 1997, A\&A, in press.

\bibitem{}{}{} Van Dokkum, P.G., and Franx, M., 1996, MNRAS, 281, 985.

\bibitem{}{}{} Worthey, G., 1994, ApJS, 95, 107.

\end{thebibliography}
\end{document}